\documentclass[epj]{svjour}
\usepackage{graphics}
\usepackage{amssymb}
\usepackage{mathtools,cuted}
\usepackage{color}
\usepackage{lipsum}
\usepackage{amsmath}
\usepackage[square, comma, sort&compress, numbers]{natbib}
\usepackage{dcolumn}% Align table columns on decimal point
\usepackage[figuresright]{rotating}
\usepackage[colorlinks=true,linkcolor=blue]{hyperref}
\begin{document}
\title{ The effects of overtaking strategy in the Nagel-Schreckenberg model}
\author{Zhu Su \and Weibing Deng \thanks{\email{wdeng@mail.ccnu.edu.cn}}\and Longfeng Zhao \and Jihui Han \and Wei Li \thanks{\email{liw@mail.ccnu.edu.cn}}\and Xu Cai% etc
%\thanks{\emph{}}%
}                     % Do not remove
%      % Insert a name or remove this line
%
\institute{Complexity Science Center, Institute of Particle Physics, Central China Normal University, Wuhan 430079, China}
\abstract{
 	Based on the Nagel-Schreckenberg (NS) model with periodic boundary conditions, we proposed the NSOS model by adding the overtaking strategy (OS). In our model, overtaking vehicles are randomly selected with probability $q$ at each time step, and the successful overtaking is determined by their velocities. We observed that (i) traffic jams still occur in the NSOS model; (ii) OS increases the traffic flow in the regime where the densities exceed the maximum flow density. We also studied the phase transition (from free flow phase to jammed phase) of the NSOS model by analyzing the overtaking success rate, order parameter, relaxation time and correlation function, respectively. It was shown that the NSOS model differs from the NS model mainly in the jammed regime, and the influence of OS on the transition density is dominated by the braking probability $p$.
 % end of PACS codes
} %end of abstract
\maketitle

\section{Introduction}
Various dynamical models \cite{Chowdhury2000,Helbing2001,Nagatani2002,Mahnke2005} have been proposed to explain the phenomena generated by traffic flow, such as fluid dynamical models, gas kinetic models and vehicle-following models \cite{Treiber1999,Thomas1993,Jamison2009,Yang2015}. From the microscopic point of view, vehicles can be represented by particles. Meanwhile, the way that vehicles influence other vehicles' movements is treated as interactions among particles. The traffic system can be regarded to be composed of interacting particles far from equilibrium. Therefore, the vehicular traffic offers the possibility to study various fundamental dynamics of non-equilibrium systems, which are of interest in statistical physics \cite{Helbing2001}.\\
\indent During the last decades, cellular automata (CA) \cite{Wolfram1983} have obtained popularity in simulating large networks, due to their simplicity and ability. One of the early CA based on traffic models is the NS model \cite{Nagel1992} developed by Nagel and Schreckenberg. Then, in order to make the CA more realistic, a large amount of modified models\cite{Chowdhury1997,Knospe1999,CHEN2001,Huang2001,Li2001,Li2006,Jiang2006,Gao2007,Jetto2010,Echab2014,Feng2015} have been proposed by changing/adding some mechanisms on the NS model. Meanwhile, many theories and methods \cite{Schreckenberg1995,Schadschneider1997,Krauss1997,Eisenblatter1998,Csanyi1999,Souza2009,Zhu2007,Zhang2014,Qiu2014} that study the properties of the NS model have been proposed. For example, the jamming transition from a free flow phase to a congested phase is analyzed in detail by studying the relaxation into the steady state, measurements of an order parameter and the investigation of correlation functions \cite{Eisenblatter1998,Csanyi1999,Souza2009}.\\ 
\indent The NS model is a probabilistic cellular automaton, in which time and velocities are discrete, and the road is divided into sites. With periodic boundary conditions, the model deals with traffic flow of $N$ vehicles moving in a one-dimensional lane of $L$ sites. Each vehicle occupies one site. Each site can be either empty or occupied. Each vehicle has an integral velocity between $0$ and $v_{max}$ (the maximum speed). Let $v(j,t)$ and $x(j,t)$ denote the velocity and position of the $j$th vehicle at time $t$, respectively. Let $d(j,t)$ be the current empty sites in front of the $j$th vehicle, $d(j,t)=x(j+1,t)-x(j,t)-1$. The evolution rules of the NS model are as follows (parallel dynamics):
\begin{description}
	\item (1) Acceleration,
	$$v(j,t_1) \rightarrow min(v(j,t)+1,v_{max});$$
	\item (2) Deterministic deceleration,
	$$ v(j,t_2) \rightarrow min(v(j,t_1),d(j,t)); $$
	\item (3) Random deceleration with probability $p$,
	$$ v(j,t_3) \rightarrow max(v(j,t_2)-1,0);$$
	\item (4) Movement,
	$$ x(j,t+1) \rightarrow x(j,t)+v(j,t+1).$$
\end{description}
\begin{figure}[h]
	\centering
	\includegraphics[scale=0.25]{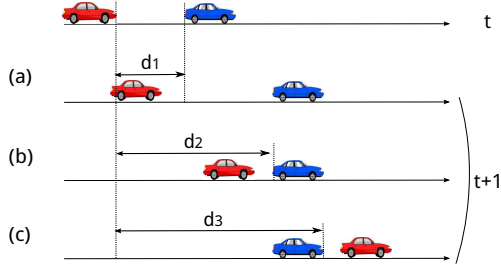}
	\caption{\label{diagram}(Color online) Illustration of updating and moving in the NS model and NSOS model, the red vehicle follows the blue one. (a) In the NS model, the red vehicle is only able to move in the regime $d_1$. In the NSOS model, if the red vehicle is an overtaking one: (b) If the value of its velocity is smaller than $d_2$, it follows the blue one. (c) If it could move more than $d_3$, it overtakes the blue one.}
\end{figure}
\begin{figure}[h]
	\centering
	\includegraphics[scale=0.3]{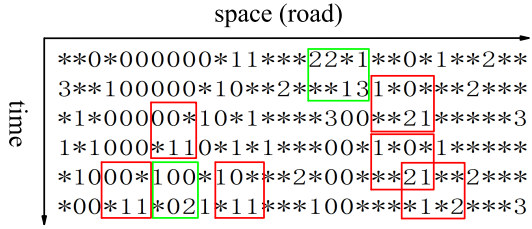}
	\caption{\label{spacetime}(Color online) Space-time diagram for $L=30$, $p=0.3$, $q=0.25$, $v_{max}=5 $ and $\rho=0.5$. The Star stands for a free site, and the number stands for the velocity of a vehicle in that site after moving. With overtaking probability $q$, overtaking vehicles (colored rectangles) try to overtake the vehicle ahead, and the overtaking vehicles in the green rectangles have succeeded while the ones in the red rectangles have failed. Jams still occur in the case of $\rho=0.5$. }
\end{figure}
\indent Here, $t_{1}$, $t_{2}$, and $t_{3}$ are simply auxiliary intervals between $t$ and $t+1$ for the convenience of implementing the CA rules computationally. The NS model is a very simple model which does not consider overtaking behavior. According to Step $(2)$, the following vehicle is only able to move in the $d(j,t)$ gap. For example, Fig.\,\ref{diagram} (a) shows the mechanism of the original NS model, where the red car is only able to move within the gap $d_1$ at time $t+1$. However, in actual traffic situation, the red car could move much further within the gap $d_2$ (Fig.\,\ref{diagram} (b)) or it could overtake the blue car, i.e., it could move further than the distance $d_3$ (Fig.\,\ref{diagram} (c)). \\
\indent Considering the above facts, we proposed the NSOS model. In the real traffic, overtaking needs to be carried out on a surpass lane. If such a lane is occupied by other vehicles and not suitable for overtaking, the overtaking vehicle will become the ordinary one. Moreover, the human behavior is complex, and the overtaking decision is made by the driver. Due to these uncertainties, we introduce the probability $q$ to describe the randomness of overtaking. In the NSOS model, each vehicle becomes an overtaking vehicle with probability $q$ at each time step. Therefore, two kinds of vehicles exist in the NSOS model: one is the overtaking vehicle, and the other is the ordinary vehicle. Their types are varied according to the probability $q$. In fact, we ignore the detailed overtaking process and assume all the vehicles move on the same lane, based on the fact that the overtaking vehicles frequently return to the original lane.\\
\indent In this paper, we analyze the effects of overtaking by discussing the relaxation time, the spatial correlation function and the order parameter. The results are compared to those of the NS model. Moreover, overtaking success rate we proposed was also studied, which can be used to depict the phase transition.\\
\indent The paper is organized as follows: In Section $2$, the NSOS model is introduced. Simulation results are discussed in Section $3$, including the space-time diagram, fundamental diagram, overtaking success rate, order parameter, relaxation time and spatial correlation. Conclusions are given in Section $4$. \\
\section{The NSOS model}
\begin{figure*}[htbp]	
	\begin{minipage}[t]{.45\textwidth}
		\centering
		\includegraphics[scale=0.25]{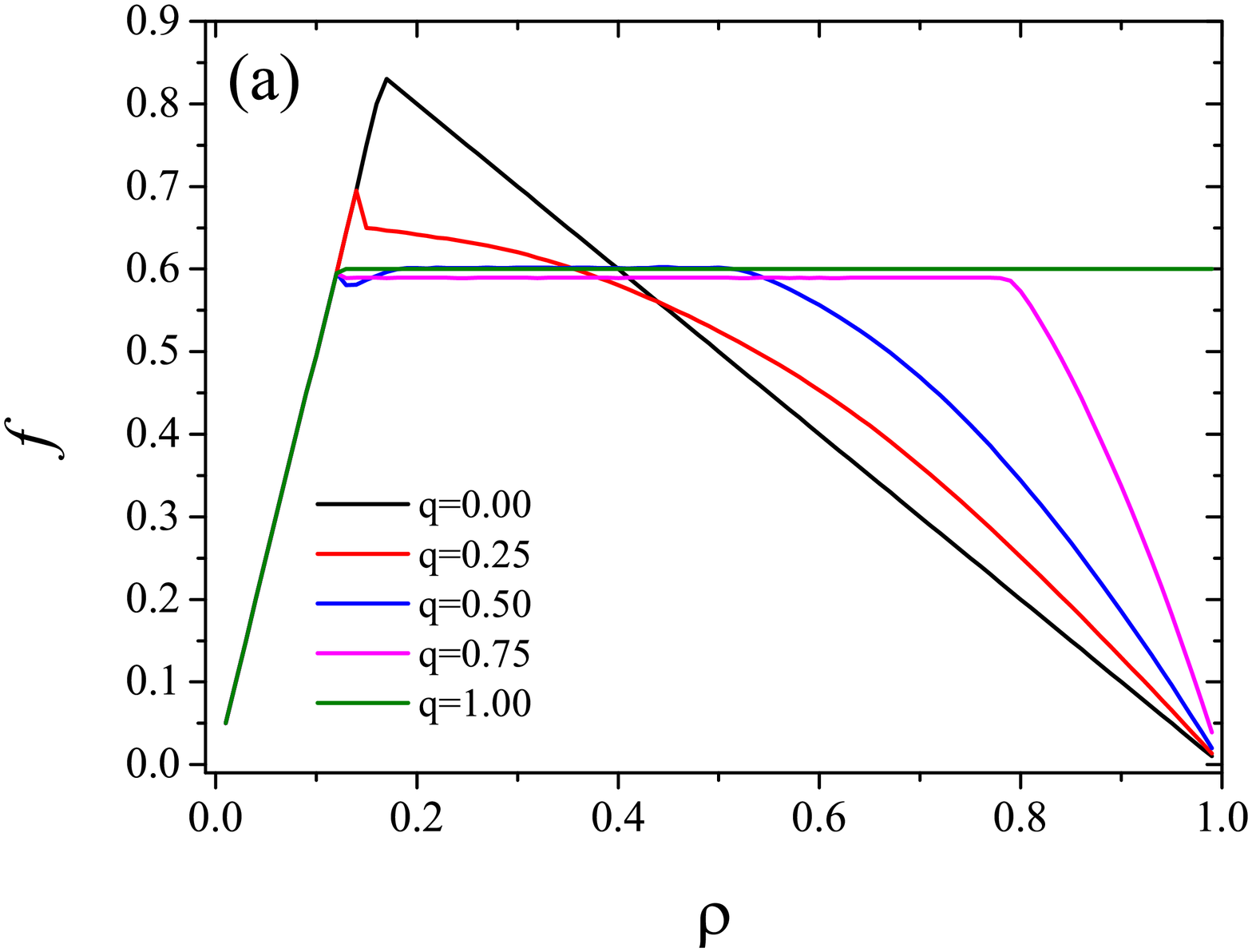}
	\end{minipage} %
	\begin{minipage}[t]{.45\textwidth}
		\centering
		\includegraphics[scale=0.25]{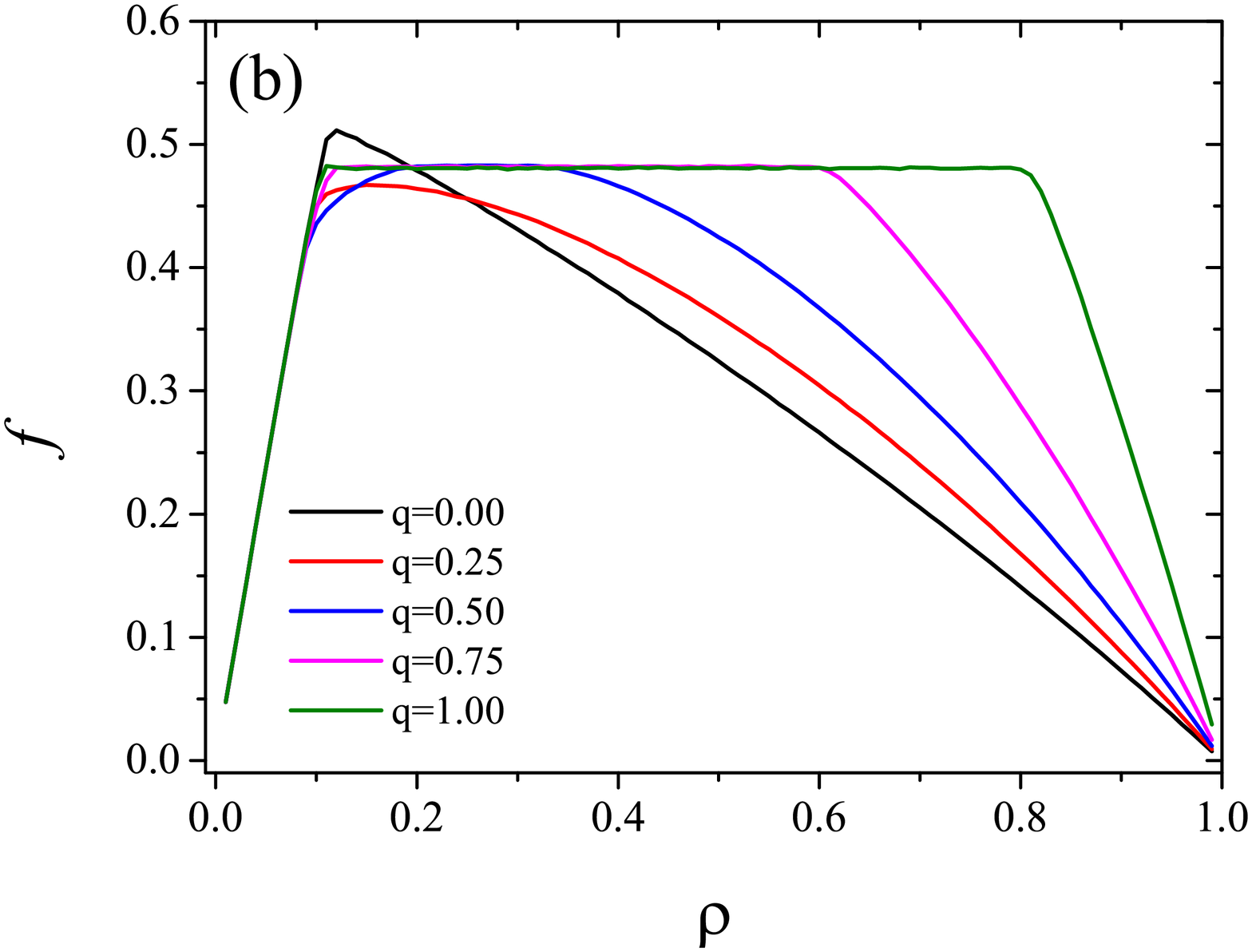}
	\end{minipage}\\
\caption{\label{flow2}(Color online) Fundamental diagrams for various values of overtaking probability $q$ in the case of two random braking probabilities: (a) $p=0$, (b) $p=0.25$ with $v_{max}=5$.}
\end{figure*}
\begin{figure*}[htbp]	
	\begin{minipage}[t]{.45\textwidth}
		\centering
		\includegraphics[scale=0.25]{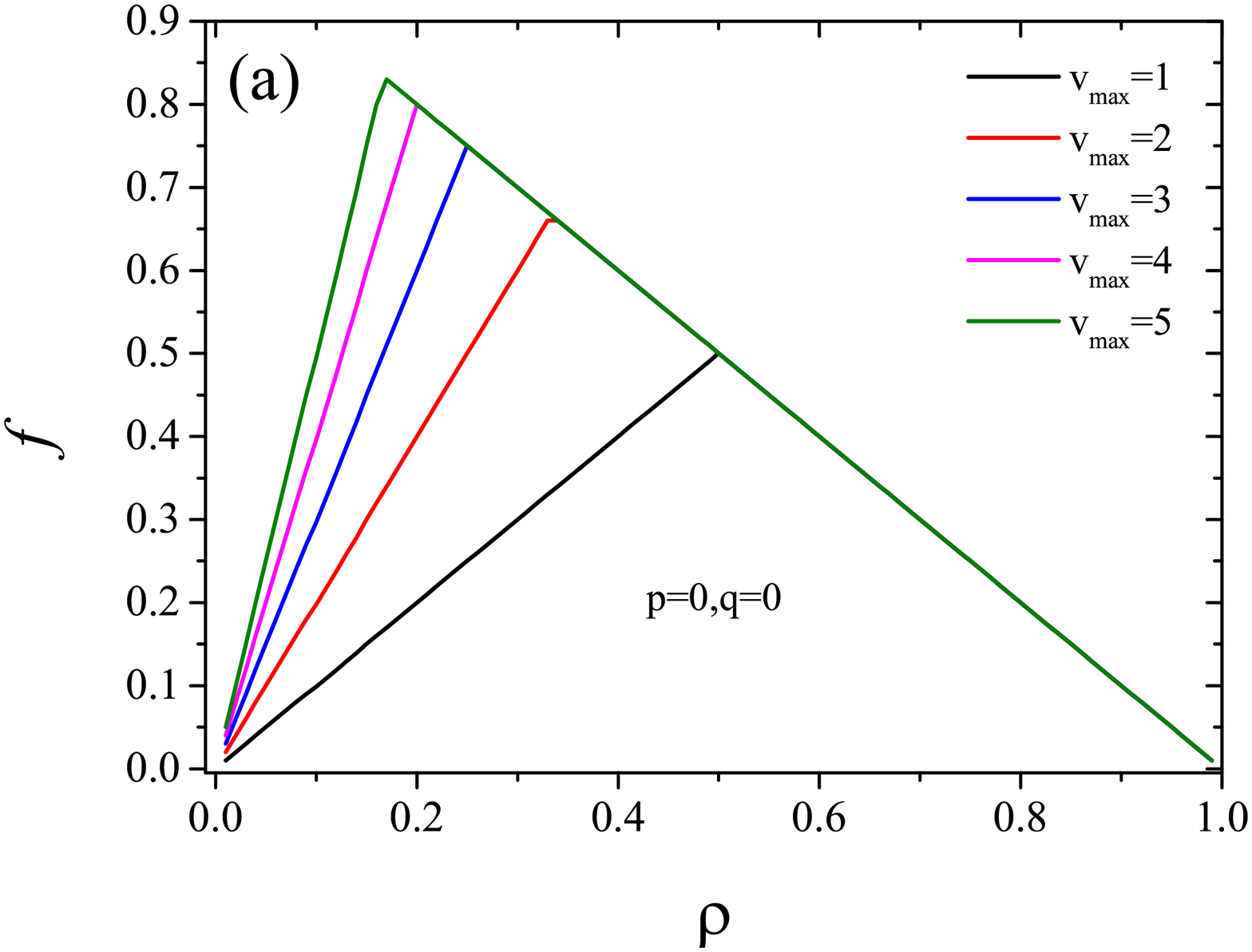}
	\end{minipage}%
	\begin{minipage}[t]{.45\textwidth}
		\centering
		\includegraphics[scale=0.25]{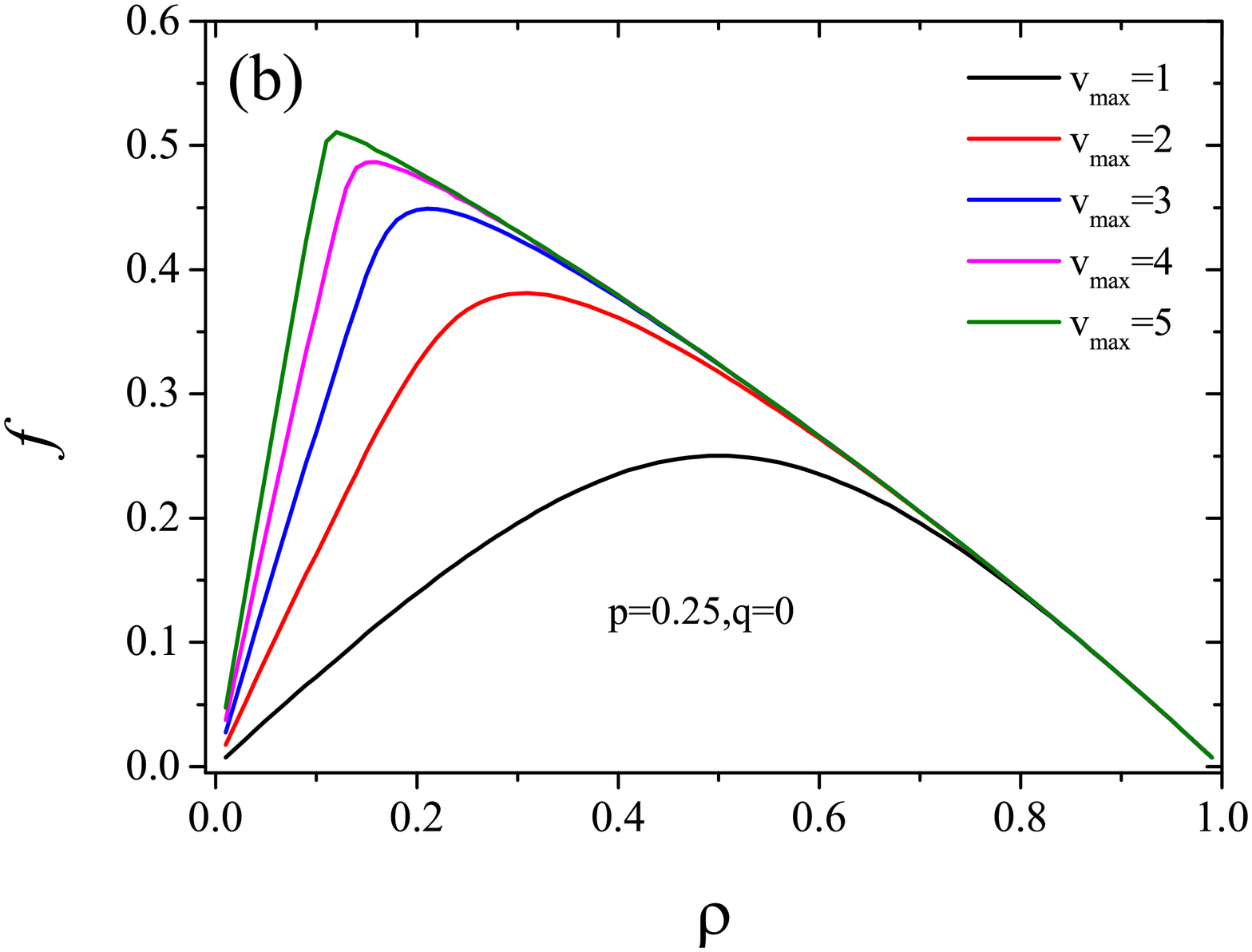}
	\end{minipage}\\
	\begin{minipage}[t]{.45\textwidth}
		\centering
		\includegraphics[scale=0.25]{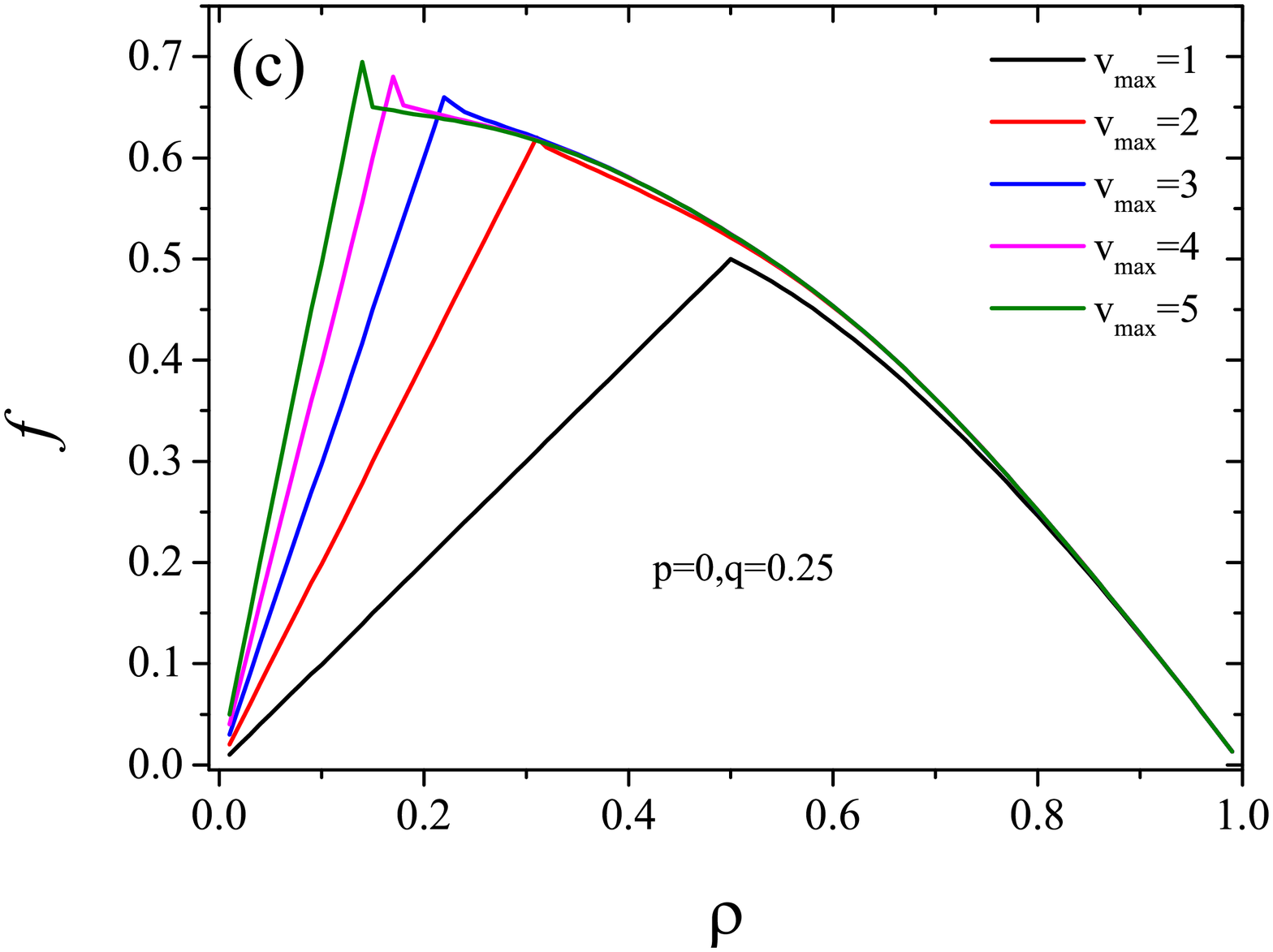}
	\end{minipage}%
	\begin{minipage}[t]{.45\textwidth}
		\centering
		\includegraphics[scale=0.25]{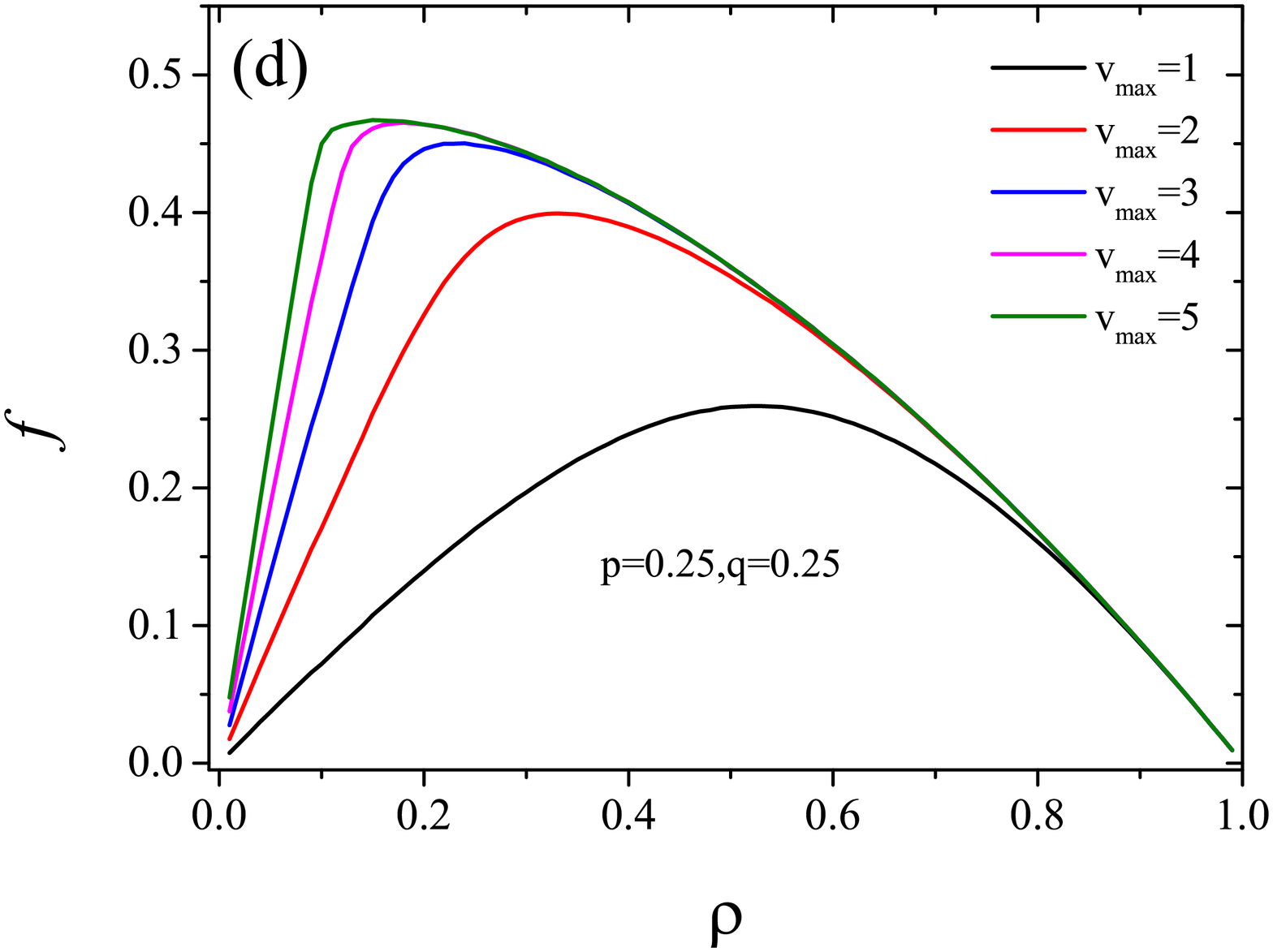}
	\end{minipage}\\
	\caption{\label{flow1}(Color online) Fundamental diagrams for various values of $v_{max}$ in different cases. (a) Deterministic NS model in the case of $p=0,q=0$. (b) NS model in the case of $p=0.25,q=0$. (c) NSOS model without random braking in the case of $p=0,q=0.25$. (d) NSOS model in the case of $p=0.25,q=0.25$.} 
\end{figure*}
The NSOS model is a modified NS model incorporating overtaking strategy. In the NSOS model, every vehicle could be an overtaking one with probability $q$ at each time step. This indicates that the system has $qN$ overtaking vehicles on average at each time step. Overtaking vehicles will try to overtake the preceding ones, but whether they succeed or not depends on their own velocities and the positions of preceding vehicles at next time step. In order to avoid collisions (vehicle reaches the same position with the preceding one at next time step), the velocity of overtaking vehicle reduces by one. For simplicity, each overtaking vehicle is only able to overtake one vehicle per time step, and its position will locate in front of its preceding vehicle if it overtakes successfully. If two consecutive vehicles are both overtaking ones, and the preceding overtakes successfully, then the following one can not overtake successfully any more. The velocities of overtaking vehicles are reduced by one with probability $p$, except for the successfully overtaking vehicles. If the vehicle is an ordinary one, we update its velocity according to the rules (1)-(3) of NS model. The detailed updating rules are as follows:
\begin{description}
	\item[1.] At time $t$, the $j$th vehicle becomes an overtaking vehicle with probability $q$, otherwise it is an ordinary one.
	\item[2.] Update the velocity:
	\begin{description}
	\item[(I)] If the $j$th vehicle is an ordinary one:
	\begin{description}
	\item[(1)] Acceleration: \\
	$v(j,t_{1}) \rightarrow min(v(j,t)+1, v_{max})$.
	\item[(2)] Deceleration: \\
	$v(j,t_{2}) \rightarrow min(v(j,t_{1}),d(j,t))$.
	\item[(3)] Random braking: \\
	$v(j,t_3) \rightarrow max(v(j,t_{2})-1,0)$ with the probability $p$.
	\end{description}
	\item[(II)] If the $j$th vehicle is an overtaking one:
	\begin{description}
	\item[(1)] Acceleration: \\
	$v(j,t_{1}) \rightarrow 	min(v(j,t)+1, v_{max})$.
	\item[(2)] If $v(j,t_{1}) > d(j,t) +v(j+1,t+1)$, the position $d(j,t)+v(j+1,t+1)+1$ is empty and the $(j+1)$th vehicle does not overtake successfully,
		\item[(i)] Overtaking: \\
		$v(j,t_{3}) \rightarrow d(j,t)+v(j+1,t+1)+1.$
	\item[(3)] Otherwise,
		\item[(i)] Deceleration: \\
		$v(j,t_{2}) \rightarrow min(d(j,t)+v(j+1,t+1)-a,v(j,t_{1}))$.
		\item[(ii)] Random braking with probability $p$: \\
		$v(j,t_{3}) \rightarrow min(v(j,t_{2})-1,0)$.
	\end{description}	
	\end{description}		
	\item[3.] Movement:\\
	 $x(j,t+1)=x(j,t)+v(j,t_{3})$.
\end{description}
\indent Here, $v(j,t)$ denotes the velocity of the $j$th vehicle at time $t$ and $x(j,t)$ denotes its corresponding position. The number of empty sites in front of the $j$th vehicle is denoted by $d(j,t)=x(j+1,t)-x(j,t)-1$. To avoid collisions, we assume $a=2$ if the $(j+1)$th overtaking vehicle overtakes successfully. In other cases, $a=1$.\\
\indent Since the velocity of overtaking vehicle at time $t+1$ is relate to its preceding vehicle's velocity, we should know the preceding vehicle's velocity at time $t+1$ first. Fortunately, the velocity of ordinary vehicle is independent of its preceding vehicle's, so we could update the ordinary vehicles' velocities first, and then update their rear overtaking vehicles' velocities. In our simulations, we use the parallel update and periodic boundary conditions. In order to ensure the ordinary vehicles always exist in the NSOS model, which makes the update of overtaking vehicles work, we let the first and last vehicles be the ordinary vehicles all the time.
\section{Simulation results}
In this section, we will analyze simulation results of the NSOS model, including the space-time diagram, fundamental diagram, overtaking success rate, order parameter, relaxation time and spatial correlation.\\
\indent In our simulations, the system size is $L=1000$ and the maximum velocity is assigned as $v_{max}=5$ unless noted otherwise. Simulations begin with vehicles randomly distributed among the system. A single simulation runs for $20000$ time steps. The sampling for the calculation of the results is done for the last $10000$ time steps and each data point is averaged over $100$ different initial configurations.
\subsection{Space-time diagram}
\begin{figure}[h]
	\begin{minipage}{18pc}
		\centering
		\includegraphics[scale=0.25]{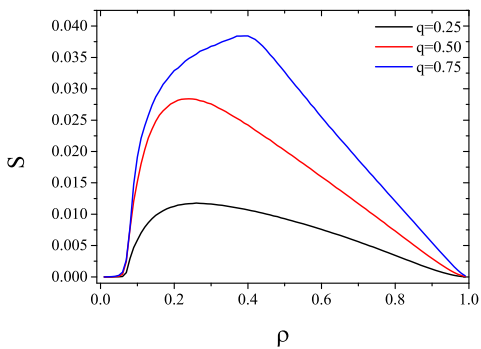}
	\end{minipage}\hspace{1pc}
	\caption{\label{success1}(Color online) The overtaking success rate $S$ as a function of density $\rho$ for the three different conditions. The parameters of the NSOS model are chosen as $v_{max}=5$ and $p=0.5$.}
\end{figure}
\begin{figure}[h]
	\begin{minipage}{18pc}
		\centering
		\includegraphics[scale=0.25]{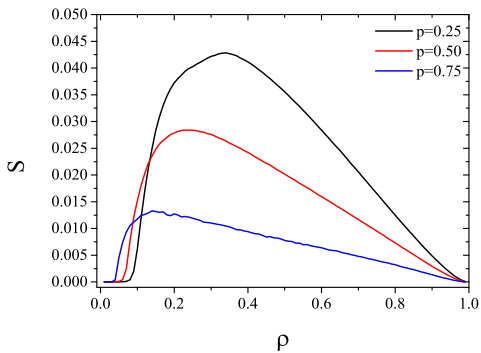}
	\end{minipage}
	\caption{\label{success2}(Color online) The overtaking success rate $S$ as a function of density $\rho$ for $v_{max}=5$, $q=0.5$ and for different values of $p$.}
\end{figure}
\indent Fig.\,\ref{spacetime} presents a snapshot of a space-time diagram of the NSOS model simulated with $L=30$, $p=0.3$, $q=0.25$, $v_{max}=5$ and $\rho=0.5$. The space direction is horizontal, the time coordinate is downward, the number stands for the vehicle's velocity after moving, and the star stands for a free site in which vehicle moves to the right. The colored rectangles contain overtaking vehicles. We can see that only a few (marked by green rectangles) overtake successfully. In the red rectangles, unlike the ordinary vehicles, the overtaking vehicles could move further to the position that was occupied by their preceding ones at the previous time step, which is a common phenomenon on real traffic. \\
\indent We can see that congestion clusters still occur, and it indicates overtaking can not avoid traffic jams. The fact that sudden jams occur is not surprising since the NSOS model has much more situations which require strong braking compared to the NS model. For example, in the top green rectangle in Fig.\,\ref{spacetime}, the speed of overtaken vehicle changes from one to zero at the following step, which will induce a jam.
\subsection{Fundamental diagram}
\indent In order to study the effects of the overtaking strategy in the NSOS model, we generate the fundamental diagrams under different traffic conditions. The average observable flow $f$ versus density $\rho$ is presented in the fundamental diagram.\\
\indent Firstly, we plot in Fig.\,\ref{flow2} the fundamental diagrams with maximum velocity $v_{max}=5$ in various situations of different overtaking probability $q$ and different random braking probability $p$. One observes that all the curves collapse in free-flow regime, since the overtaking vehicles make no contributions to the flow due to no interactions between them. It is interesting to notice that the values of maximum flow are suppressed for non-vanishing $q$. The major reason is that the overtaking mechanism makes more vehicles move closer, which will lead to more emergency brakes. Unlike the NS model, it is found that there exists a plateau at the large value of $q$. This is because the larger the overtaking probability, the more overtaking vehicles exist in the system that move synchronously without interacting with each other. Moreover, unlike in Fig.\,\ref{flow2} (b) where $q=0.25$, the plateau never decays for $p=1$ in Fig.\,\ref{flow2} (a) where $q=0$, as the absence of random braking. Another phenomenon is the flow enlarges as $q$ increases in the jammed regime, which means that overtaking strategy increases the traffic flow in the jammed therein. \\ 
\indent Fig.\,\ref{flow1} shows the fundamental diagrams for various values of $v_{max}$ in different cases. Fig.\,\ref{flow1} (a)-(d) are the deterministic NS model ($p=0,q=0$), the standard NS model ($p=0.25,q=0$), the deterministic NSOS model($p=0,q=0.25$), and the NSOS model ($p=0.25,q=0.25$), respectively. Firstly we can see that the density corresponding to the maximum flow $\rho(f_{max})$ is apparently declined with the increase of $v_{max}$. In fact, in comparison with Fig.\,\ref{flow1} (a) and (b), as $p$ increases, $\rho(f_{max})$ is also declined. While the value of $\rho(f_{max})$ does not change even when $q$ is changed. Therefore, the main factors that determine the values of $\rho(f_{max})$ are the maximum velocity $v_{max}$ and the braking probability $p$, while the overtaking probability $q$ just enlarges the flow in the jammed regime.\\
\indent Specifically, according to \cite{Eisenblatter1998}, only in the deterministic NS model (Fig.\,\ref{flow1} (a)), the flow is solved exactly given by $f(\rho)=min(\rho v_{max},1-\rho)$, and $\rho(f_{max})$ coincides with the critical density $\rho_c$, which is the transition density from the free flow phase to the jammed phase. In other words, $\rho(f_{max})$ is not equal to $\rho_c$ except in the deterministic NS model. 
\subsection{Overtaking success rate}
Since not all the overtaking vehicles can overtake successfully in the NSOS model, it is natural to explore the overtaking success rate. Here, an overtaking vehicle is marked by $s$, and $s=1$ if it overtakes successfully, otherwise $s=0$. \\
\indent The overtaking success rate per time step per vehicle is defined by 
\begin{equation}
	S=\frac{1}{T}\frac{1}{N_{0}}\sum\limits_{t=t_{0}+1}^{t_{0}+T}\sum\limits_{i=1}^{N_{0}}s_{i},
	\label{r}
\end{equation}
where $N_{0}$ is the number of overtaking vehicles in the system, $t_{0}$ is the time sufficiently long after the system has reached a steady state, taken as $t_{0}=10^{5}$. \\
\indent In fact, overtaking success rate $S$ is a function of the density $\rho$, braking probability $p$ and overtaking probability $q$. First, we show in Fig.\,\ref{success1} the overtaking success rate $S$ as a function of density $\rho$ calculated under different overtaking probabilities $q$. The simulation parameters are $v_{max}=5$ and $p=0.5$. At low densities, overtaking will not occur until the density reaches a critical density $\rho_{c}$. This is because in the free-flow region ($\rho \leq \rho_{c} $) all vehicles move with the velocity $v_{max}$. The average gap between vehicles is superior to the maximum speed $v_{max}$ and thus successful overtaking vehicles do not exist.\\
\indent From Fig.\,\ref{success1}, we observe that the values of the critical densities $\rho_{c}$ almost remain the same. This is because braking probability $p$, which essentially determines the critical densities shown in Fig.\,\ref{success2}, is the same in the three different conditions.\\
\indent Above the critical density, $S$ increases with the increase of density, and reaches a maximum value, then decreases with much larger density. Comparing the three different conditions, we can see that the larger overtaking probability leads to larger value of $S$. At relatively high density, the overtaking success rate $S$ reaches the maximum value. From our daily traffic experiences on highways, overtaking usually occurs when the vehicles are partially in jams and free flow. \\
\indent To show how $S$ depends on the braking probability $p$, we plot in Fig.\,\ref{success2} the values of $S$ against density $\rho$ with maximum velocity $v_{max}=5$, overtaking probability $q=0.5$ and for different values of $p$. We observe that the critical density $\rho_{c}$ decreases when increasing $p$. Another interesting result is that beyond $\rho_{c}$, the overtaking success rate $S$ increases when $p$ decreases. These results could be explained by the fact that when the braking probability $p$ increases, the vehicles' movements become slow. In a certain way, the overtaking is more difficult.
\begin{figure}[h]
	\centering
	\includegraphics[scale=0.25]{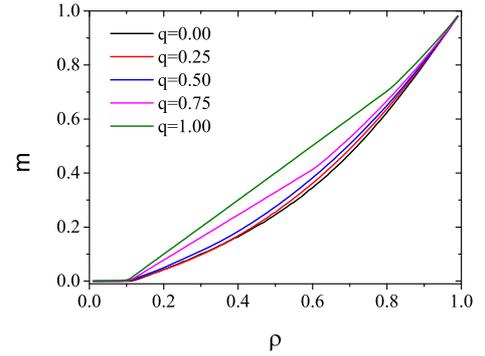}
	\caption{\label{order1}(Color online) Order parameter $m$ vs.\,density $\rho$ ($L=1000$) for various values of $q$ in the case of $p=0.25$. They do not vanish exactly for $\rho <\rho_{c}$ but converges smoothly to zero.}
\end{figure} 
\begin{figure}[h]
	\centering
	\includegraphics[scale=0.25]{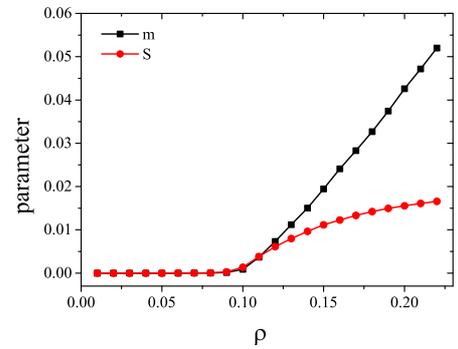}
	\caption{\label{order2}(Color online) The behaviors of order parameter $m$ and overtaking success rate $S$ near the critical density. Here, $p=0.25$ and $q=0.25$.}
\end{figure} 
\begin{figure}[h]
	\centering
	\includegraphics[scale=0.25]{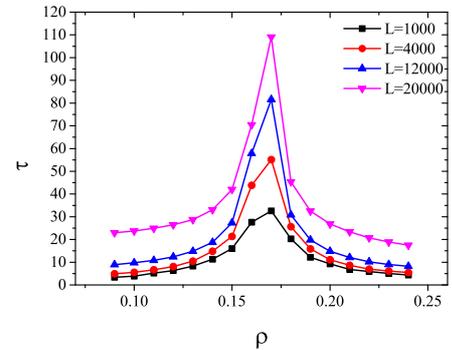}
	\caption{\label{relaxation1}(Color online) Relaxation time near the transition density for different system sizes for the deterministic case ($p=0,q=0$).}
\end{figure}
\begin{figure}[h]
	\centering
	\includegraphics[scale=0.25]{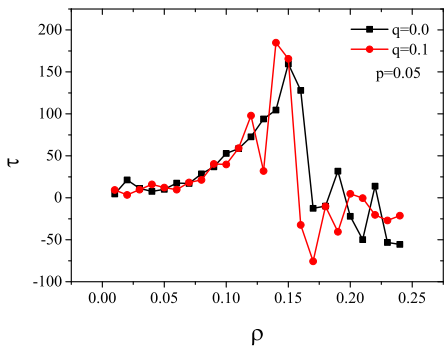}
	\caption{\label{relaxation2}(Color online) Relaxation time near the transition density in the NSOS model and NS model under $p=0.05$.}
\end{figure}
\subsection{Order parameter}
In the NS model, the phase transition, from the free flow regime to the jammed phase, has been investigated \cite{Eisenblatter1998,Csanyi1999,Souza2009}. Here, we use the same approaches to analyze the effects of overtaking strategy in the NSOS model.\\
\indent To look for the transition between the free and jammed phase, an order parameter has been introduced in \cite{Eisenblatter1998}, which is defined by 
\begin{equation}
	m=\frac{1}{T}\frac{1}{L}\sum\limits_{t=t_{0}+1}^{t_{0}+T}\sum\limits_{i=1}^{L}n_{i}n_{i+1},
	\label{m}
\end{equation}
with $n_{i}=0 $ if a site is empty, and $n_{i}=1$ for a site occupied by a vehicle. In fact, the order parameter $m$ is counting occurrence of neighboring vehicles, which describes the jamming level.\\
\indent In the determined NS model ($p=0$), the transition density is $\rho_{c}=1/(v_{max}+1)$. Below the transition density, the values of order parameter are zero because every vehicle moves with $v_{max}$. Within the jammed phase, the flow is limited by the number of empty sites and motionless vehicles occur.\\
\indent When $p>0$, the behavior of the order parameter qualitatively changes in the vicinity of the transition density in the NS model. It does not vanish exactly at $\rho < \rho_{c}$ but converges smoothly to zero even for small values of the braking probability $p$.\\
\indent In order to investigate the effects of overtaking in the NSOS model, we plot in Fig.\,\ref{order1} the values of $m$ against the vehicle density $\rho$ with maximum velocity $v_{max}=5$, braking probability $p=0.25$ and for various values of $q$. We find that the order parameters in the NSOS model have the similar behavior in the NS model, which belongs to the continuous phase transition. The disparity is that the values of order parameters are slightly increased with the growth of the values of $q$ in the jammed phase. This result could be explained by the fact that the NSOS model provides much more configurations of adjacent vehicles than the NS model, which leads to the increase of order parameter $m$. Again, we also observe that the critical density remains unchanged with respect to the different overtaking probabilities.\\
\indent Fig.\,\ref{order2} shows the behaviors of order parameter $m$ and overtaking success rate $S$ near the critical density. One can see that around the critical density the order parameters are the same as the overtaking success rate $S$, which further confirms that the overtaking success rate is able to identify the critical state. In fact, below the critical density, there is no difference between the order parameter and overtaking success rate, while the disparity occurs above the critical density.
\subsection{Relaxation time}
\begin{figure}[h]
	\centering
	\includegraphics[scale=0.25]{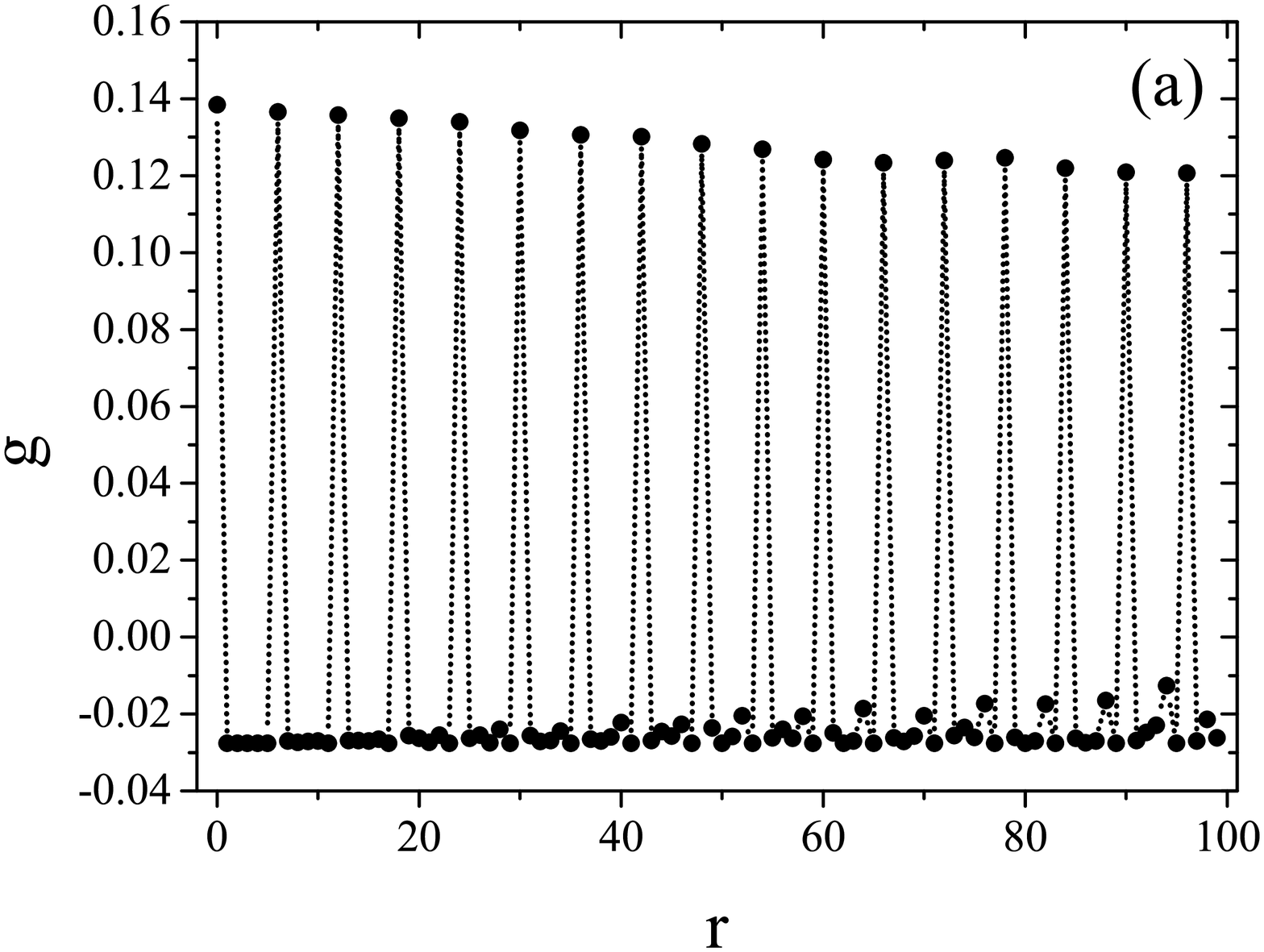}
	\centering
	\includegraphics[scale=0.25]{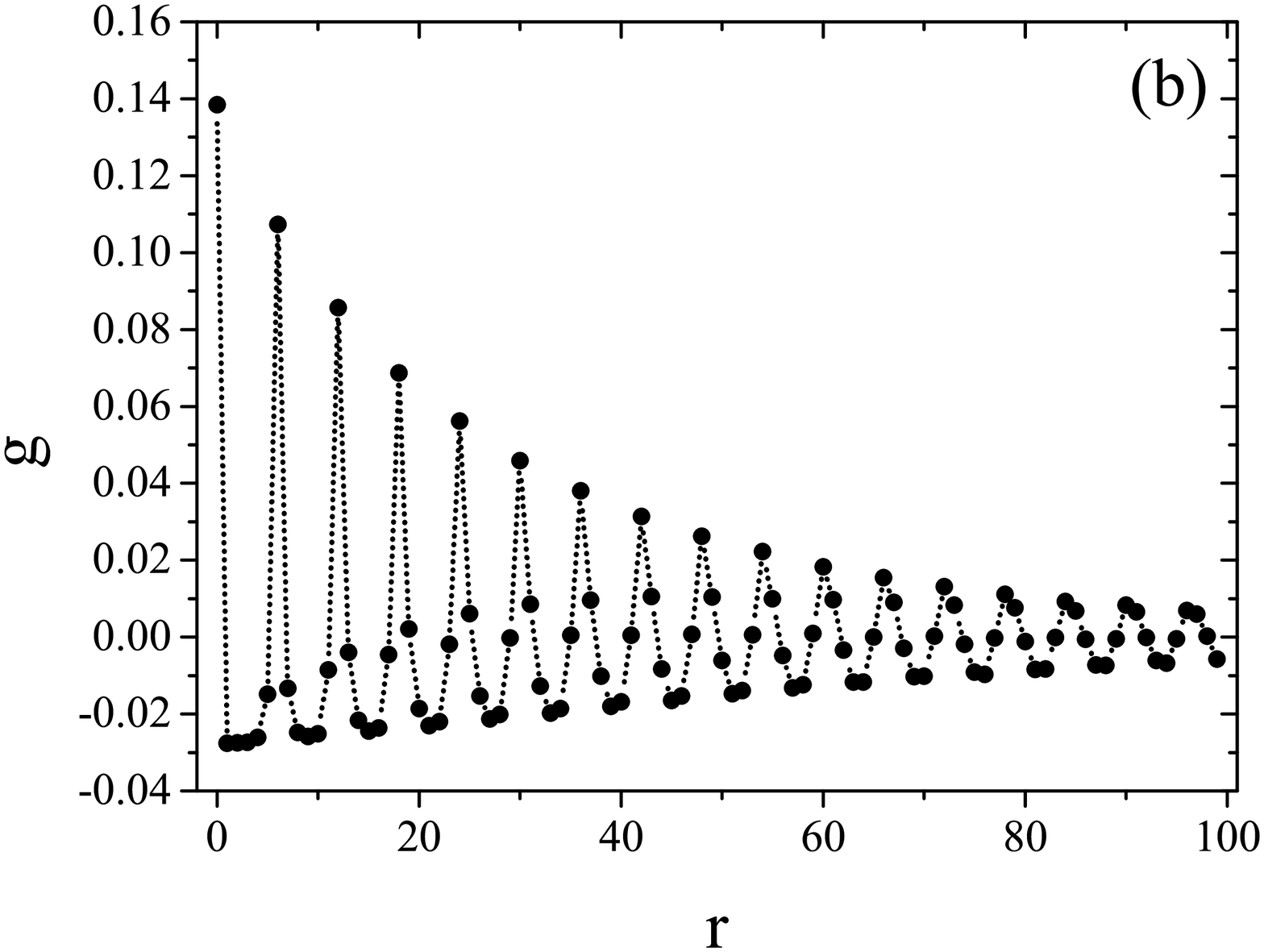}
	\centering
	\includegraphics[scale=0.25]{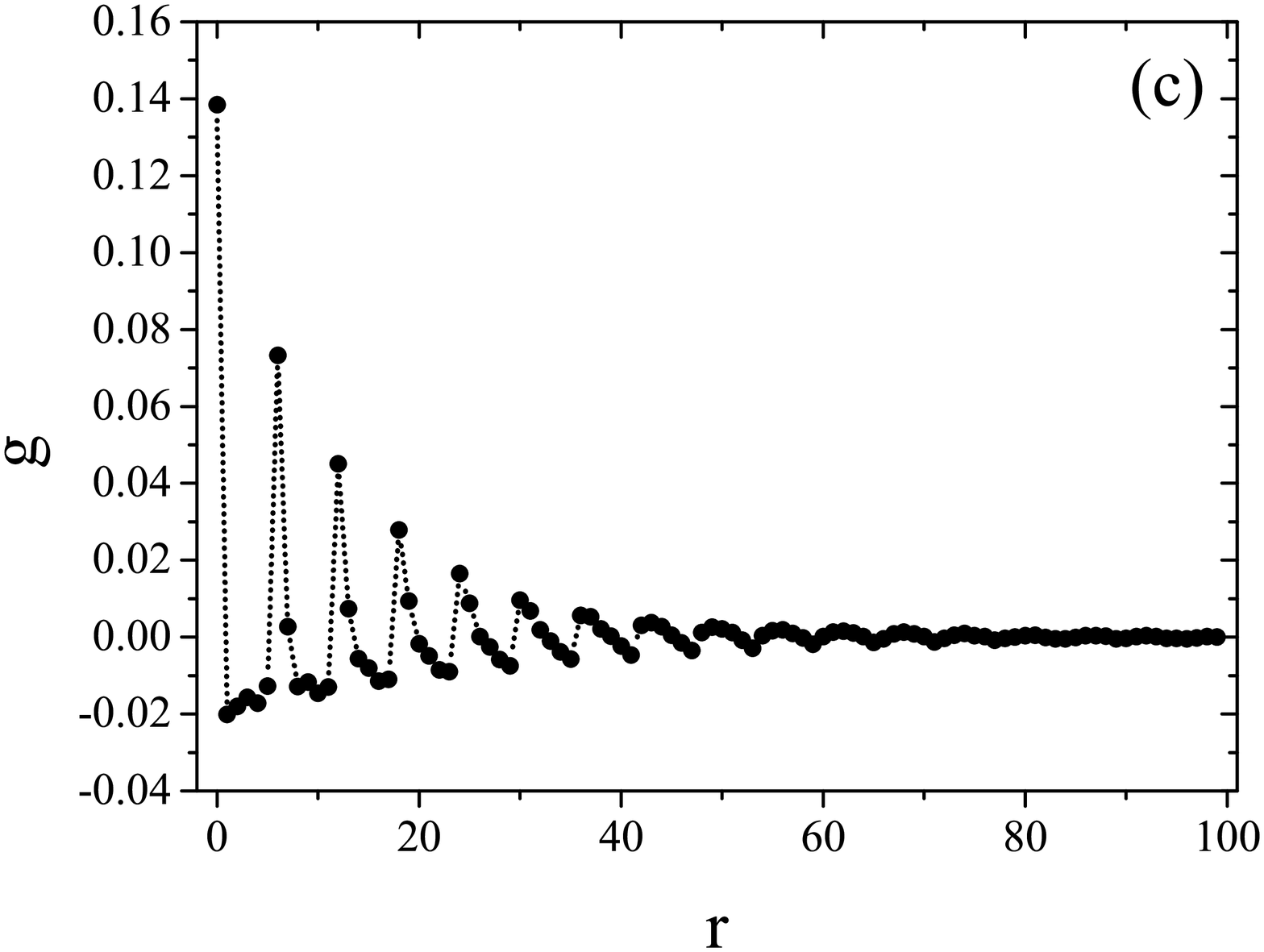}
	\caption{\label{correlation1}(Color online) (a) Correlation function in the vicinity of the phase transition for the deterministic case. At $\rho=\rho_{c}$ the amplitude is independent of the distance $r$. (b) In the case of $q=0$, the amplitude of the correlation function decays exponentially. (c) In the case of $q=0.1$, the amplitude of the correlation function decays exponentially faster compared with (b).} 
\end{figure}
\indent A characteristic feature of a second order phase transition is the divergence of the relaxation time at the transition point. According to Ref. \cite{Csanyi1999}, the relaxation time is characterized by the parameter (Due to the dimensionless units no normalization is introduced)
\begin{equation}
	\tau=\int_{0}^{\infty}[min\{v^{*}(t),\langle{\overline{v}_{\infty}}\rangle\}-\langle{\overline{v}(t)}\rangle]dt,
	\label{2}
\end{equation}
where $\overline{v}(t)$ is the average velocity of all vehicles as a function of time $t$ starting from randomly positioned vehicles with zero initial velocity ($\overline{v}(t=0)=0$). For $t\rightarrow\infty$ the system reaches a stationary state with average velocity $\langle\overline{v}_{\infty}\rangle$. $v^{*}(t)$ denotes the average velocity in the acceleration phase $t\rightarrow 0$ for low vehicle density $\rho \rightarrow 0$, $v^{*}(t)=(1-p)t$ holds in this regime since the vehicles don't interact with each other. Thus the relaxation time is obtained by summing up the deviations of the average velocity $\langle{\overline{v}(t)}\rangle$ from the values of a system with one single vehicle that can move without interactions with other vehicles ($\rho \rightarrow 0$). In our simulations, the total running time of system is $10L + T$. We calculate the velocities when $t<10L$. After $10L$ time steps, the system reaches the stationary state, and then we average the stationary velocity $\overline{v}_{\infty}$ during the $T$ time steps, with $T=5000$ here. Each data point is averaged over $100$ different initial configurations.\\
\indent At first, we have reviewed the scaling behavior of the relaxation time in the deterministic model shown in Fig.\,\ref{relaxation1}. We consider different system sizes up to $L=20000$ where the position of the critical density $\rho_c$ becomes size independent. It is found that there exists a maximum value of relaxation time at the critical density. Taking the magnitude of $\tau$ as a characteristic value for the relaxation time, we can estimate the dynamical exponent. Moreover the scaling behavior of the width $\sigma(L)$ and height $\tau_{m}(L)$ of the peak has the following form,
\begin{equation}
	\tau_{m}(L) \propto L^{z},
	\sigma(L) \propto L^{-1/ \mu}.
	\label{t}
\end{equation}
We obtain the critical exponents for the deterministic case ($p=0,q=0$) are $z=0.53$ and $\mu=6.7$.\\ 
\indent In order to study the effect of overtaking on the relaxation time $\tau$, we assign $p=0.05$ and $q=0.1$ in the NSOS model, and make a comparison with the NS model ($q=0$). The results are shown in Fig.\,\ref{relaxation2}. Unlike the deterministic case, the shape is not symmetric. We find that the relaxation time of the NSOS model ($q=0.1$) is larger than that of the NS model ($q=0$) at critical density, which implies that the system will take more time to reach the stable state due to the extra fluctuations induced by the overtaking probability $q$. Another phenomenon is that the negative relaxation times occur both in the two models, this is because the system gets temporarily into states that have a higher average velocity than the stationary state, i.e., $\langle \overline{v}(t) \rangle > \langle \overline{v}_{\infty} \rangle$, in the case of $\rho> \rho_{c}$ \cite{Eisenblatter1998}. During the first few time steps, small clusters occur in the initial configuration, and then more and more vehicles get trapped into large jams. Therefore, the average flow decreases to its stationary value.
\subsection{Spatial correlation}
\begin{figure}[h]
	\centering
	\includegraphics[scale=0.25]{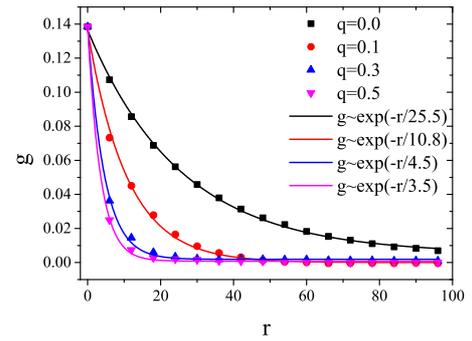}
	\caption{\label{correlation2}(Color online) Correlation length according to the maximum amplitude of correlation function in different $q$ in the case of $p=0.008$ at critical density $\rho_c=0.1666$. The dots represent the results of simulations, and the solid lines are exponential fittings. The fitting parameters are the corresponding correlation length. For the NS model ($q=0$), its correlation length is $\xi=25.5$. For the NSOS model in the case of $q=0.1$, $q=0.3$ and $q=0.5$, respectively, their correlation lengths are $\xi=10.8$, $4.5$ and $3.5$.} 
\end{figure} 
\begin{figure}[h]
	\centering
	\includegraphics[scale=0.25]{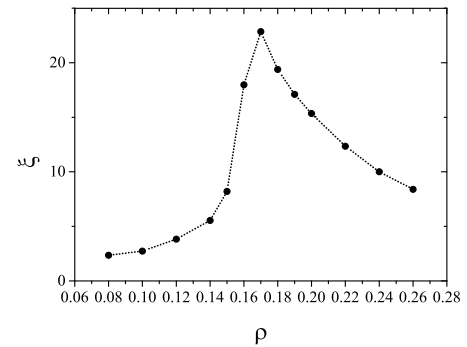}
	\caption{\label{corrlation3}(Color online) Density dependence of the correlation length in the vicinity of the phase transition density with $p=0.008,q=0$.}
\end{figure}
\begin{figure}[h]
	\centering
	\includegraphics[scale=0.25]{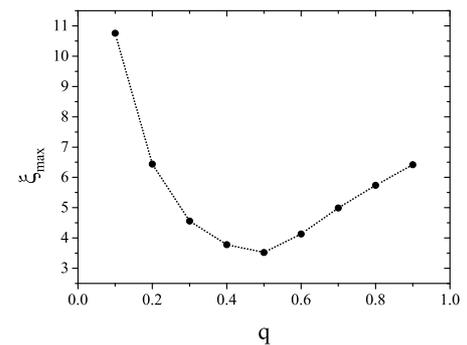}
	\caption{\label{correlation4}(Color online) The maximum correlation length $\xi_{max}$ as a function of the overtaking probability $q$ for $p=0.008$.} 
\end{figure}
\indent Another feature of second order phase transitions is the occurrence of a divergence length scale at critical density and a corresponding algebraic decay of the correlation function. The correlation function is defined as
\begin{equation}
	g(r)=\frac{1}{T}\frac{1}{L}\sum\limits_{t=t_{0}+1}^{t_{0}+T}\sum\limits_{i=1}^{L}n_{i}n_{i+r}-\rho^{2},
	\label{g}
\end{equation}
where $n_{i}=0 $ if the lattice is empty, otherwise $n_{i}=1$. Here, we obtain the same result in the vicinity of the phase transition for the deterministic NS model ($p=0,q=0$) shown in Fig.\,\ref{correlation1} (a). There is a decay of the amplitude of $g(r)$ for larger value of the distance between the sites. Precisely at $\rho_{c}$, the correlation function is solved exactly given by $g(r)=\rho_{c}-\rho_{c}^{2}$ for $r=0$ mod($v_{max}+1$) or $g(r)=-\rho_{c}^{2}$ for otherwise.\\
\indent In order to study the effect of overtaking on the correlation function, we have compared the results of the NS model with NSOS model at small value of $p$, here $p=0.008$, shown in Fig.\,\ref{correlation1} (b) and Fig.\,\ref{correlation1} (c). The common feature is that their amplitudes of the correlation function decay exponentially, except that in the NSOS model such a decay is much faster. The reason is that in the NS model after the transient state, the vehicles arrange themselves in regular intervals and therefore the ordering is long-range. In the NSOS model, with random braking and/or random overtaking, this long-range ordering is destroyed.\\
\indent Furthermore, we calculate the correlation lengths in the two models according to their maximum amplitude shown in Fig.\,\ref{correlation2}. For the NS model ($q=0$), the correlation length is $\xi=25.5$. While in the NSOS model, the correlation length is shorter, for example, $\xi$ equals $10.8$ when $q=0.1$. These results demonstrate that the overtaking reduces the spatial correlation. The maximum value of correlation length $\xi_{max}$ determines the critical density for small value of $p$ in the NS model. For example, Fig.\,\ref{corrlation3} shows the relationship between correlation against density in the case of $p=0.008$ and $q=0$. We could observe that the correlation length reaches the maximum value at the critical density. We also calculate $\xi_{max}$ for different overtaking probabilities $q$ shown in Fig.\,\ref{correlation4}. We can see that $\xi_{max}$ does not always decay with the growth of overtaking probability $q$, and there is a critical value of $q$ corresponding to the minimum correlation length. Here, in the case of $p=0.008$, the critical overtaking probability is about $q_{c}=0.5$, and the minimum correlation length is $\xi_{max}=3.5$.
\section{Conclusions}
\indent This paper has proposed the NSOS model for a simplified traffic situation. Based on the NS model, we add the mechanism of overtaking strategy by bringing in the overtaking probability $q$. Basically, one major difference between the two models is that our model makes the overtaking possible, allowing one vehicle overtakes the preceding one or moves further within larger gap.\\
\indent We have studied the properties of NSOS model under periodic boundary conditions, and made a comparison with the results of NS model. The space-time diagram tells us that the traffic jams are still spontaneous formation, even if the existence of overtaking vehicles. And we find traffic flow enlarged in the jammed regime, according to the comparison of the fundamental diagrams with that of NS model. Furthermore, we redo many phase transition calculations within the NS framework, such as the order parameter, relaxation time and spatial correlation. It is found that the critical density is dominated by the braking probability $p$ other than overtaking probability $q$. According to our model, we proposed the concept of overtaking success rate $S$. We find it could be used to characterize the phase transition, since successful overtaking vehicles do not occur in the free flow phase. \\

\section*{Author contribution statement}
X.C. and W.L. proposed the research. Z.S. and J.H. designed research, which was carried out by Z.S. Z.S., W.D. and L.Z. analyzed the results. All authors wrote, reviewed and approved the manuscript. 
\begin{acknowledgement}
This work was supported by National Natural Science Foundation of China (Grant No. 11505071), the Program of Introducing Talents of Discipline to Universities under grant No. B08033. All authors contributed equally to the paper.
\end{acknowledgement}
\bibliographystyle{elsarticle-num}
\bibliography{reference.bib}

\end{document}